\documentclass{llncs}

\usepackage{makeidx}  
\usepackage{url}
\usepackage[hidelinks]{hyperref}
\usepackage{alltt}

\makeatletter
\g@addto@macro{\UrlBreaks}{\UrlOrds}
\makeatother
\usepackage{filecontents}

\begin{document}

\title{Restpark: Minimal RESTful API\\ for Retrieving RDF Triples}

\author{Luca Matteis}
\institute{
\email{lmatteis@gmail.com}
}

\maketitle              

\begin{abstract}
How do RDF datasets currently get published on the Web? They are either available as large RDF files, which need to be downloaded and processed locally, or they exist behind complex SPARQL endpoints. By providing a RESTful API that can access triple data, we allow users to query a dataset through a simple interface based on just a couple of HTTP parameters. If RDF resources were published this way we could quickly build applications that depend on these datasets, without having to download and process them locally. This is what Restpark is: a set of HTTP GET parameters that servers need to handle, and respond with JSON-LD.
\end{abstract}

\section{Why?}
SPARQL is quite complex, and to implement it on top of existing storage solutions you need to use a proper triple store (RDF database such as Virtuoso), and be sure it implements all of SPARQL features correctly \cite{sparql}.

Having a lightweight alternative to SPARQL which developers could implement themselves on top of their existing storage solution (MySQL, MongoDB, etc.) using the language they want (PHP, Java, Node.js), would lower the entry barrier for Semantic Web data understanding and retrieval.

\section{Query parameters}

\begin{center}
\begin{tabular}{ l | l | l }
\textbf{Parameter} & \textbf{Value}	& \textbf{Description} \\
\hline
subject	& URI of subject	& Used to match a subject\\
predicate &	URI of predicate	& Used to match a predicate\\
object	& URI of node or literal	& Used to match an node or a string\\
\end{tabular}
\end{center}

\noindent Therefore a Restpark compatible query would look like:

\begin{verbatim}
/restpark?subject={subject}&predicate={predicate}&object={object}
\end{verbatim}

\section{Examples}

\textit{Get the birthdays (DBpedia) of people who acted in Star Trek (LinkedMDB):}\\[2ex]

\noindent 1) Let's first query LinkedMDB to get all the names of the actors of Star Trek:

\begin{verbatim}http://data.linkedmdb.org/restpark
    ?subject=http://data.linkedmdb.org/resource/film/675\end{verbatim}
    
\noindent This will return a JSON that we can parse and extract the actor name.\\[2ex]

\noindent 2) Now let's query DBpedia to get all the actors information (from the actor name we just retrieved above):

\begin{verbatim}http://dbpedia.org/restpark?object="William Shatner"\end{verbatim}

\noindent In this example you can see how the link (or foreign key) between the two datasets is the actor name, a literal. This method is not very accurate for linking two (or more) datasets: the literal might be slightly different between the two datasets resulting in inaccurate joins of data. 

The best way would be for LinkedMDB to provide a link (URI) to DBpedia. In the next example we will use owl:sameAs to illustrate how more accurate links can be made between various datasets.\\[2ex]

\noindent \textit{Get me the papers (DBLP) of Fellows of the British Computer Society (DBpedia):}\\[2ex]
\noindent 1) Let's first get all the Fellows from DBpedia:

\begin{verbatim}http://dbpedia.org/restpark
    ?predicate=http://purl.org/dc/terms/subject
    &object=http://dbpedia.org/resource/Category:Fellows_of_the_
    British_Computer_Society\end{verbatim}
    
\noindent 2) Now that we have a list of all the Fellows from DBpedia, we want to know if these Fellows have any external linkages. We essentially need to find out if they have an owl:sameAs predicate. For each of the Fellow returned above, we run this query:

\begin{verbatim}http://dbpedia.org/restpark
    ?subject=http://dbpedia.org/page/Tim_Berners-Lee
    &predicate=http://www.w3.org/2002/07/owl#sameAs\end{verbatim}
    
\noindent We now have a list of URIs that represent external resources. Given the initial question, we are specifically interested in resources from the DBLP dataset. So we can simply filter the result and only use URIs that are from DBLP.\\[2ex]

\noindent 3) We can finally run another query on the DBLP dataset, since we have the URI, to extract the paper information of the Fellow we retrieved in the earlier step:

\begin{verbatim}http://www4.wiwiss.fu-berlin.de/dblp/restpark
    ?subject=http://www4.wiwiss.fu-berlin.de/dblp/Tim_Berners-Lee\end{verbatim}

\bibliographystyle{unsrt}

\begin{thebibliography}{[MT1]}
\bibitem[1]{sparql} 
Buil-Aranda, Carlos, Aidan Hogan, J\"urgen Umbrich, and Pierre-Yves Vandenbussche. ``SPARQL Web-Querying Infrastructure: Ready for Action?.'' In The Semantic Web, ISWC 2013, pp. 277-293. Springer Berlin Heidelberg, 2013.
\end{thebibliography}

\end{document}